\documentclass{article}

\usepackage{graphicx}  
\usepackage{amsmath}   
\usepackage{amssymb}   
\usepackage{bm}

\usepackage[compress]{cite}

\usepackage{color}

\def\eq#1{{Eq.~(\ref{#1})}}

\def\bk#1#2#3{{\langle #1|#2|#3\rangle}}  
\def\amp#1#2{\langle #1 | #2\rangle}      

\def\bp#1{{\bm{#1}_\perp}}

  \title{Geodesic distance: A descriptor of geometry \textit{and} correlator of pre-geometric density of  spacetime events}
  \author{T. Padmanabhan\\
  IUCAA, Pune University Campus,\\
  Ganeshkhind, Pune- 411 007.\\
  {\small {email: paddy@iucaa.in}}
  }

  \date{}  

\begin{document}
  
  \maketitle

  \begin{abstract}
  Classical geometry can be described either in terms of a metric tensor $g_{ab}(x)$ or in terms of the geodesic distance $\sigma^2(x,x')$. Recent work, however, has shown that the geodesic distance is better suited to describe the quantum structure of spacetime. This is because one can incorporate some of the key quantum effects by replacing $\sigma^2$ by another function $S[\sigma^2]$ such that $S[0]=L_0^2$ is non-zero. This allows one to introduce a zero-point-length in the spacetime. I show that the geodesic distance can be an emergent construct, arising in the form of a correlator $S[\sigma^2(x,y)]=\langle J(x)J(y)\rangle$, of a pregeometric variable $J(x)$, which can be interpreted as the quantum density of spacetime events. This approach also shows why null surfaces play a special role in the interface of quantum theory and gravity. I describe several technical and conceptual aspects of this construction and discuss some of its implications.
  \end{abstract}

  \section{Geodesic distance: A replacement for the metric}
  
  The geometry of spacetime (or space, since I will work with an \textit{Euclidean} manifold in this section) is conventionally described in terms of a metric $g_{ab}(x)$ which is a local, second rank symmetric tensor. The distance between two infinitesimally separated events (or points) is then given by $ds^2 = g_{ab}dx^a dx^b$. All other geometrical features of the space(time) are then related to the metric. 
  
  There is, however, another way of describing the geometry which is conceptually far superior. This is in terms of  the geodesic distance $\sigma^2(x,x')$ [also called Synge's world function \cite{synge}; but I will use the terminology `geodesic distance'] which is related to the metric tensor by  two equations.
  The first one is:
 \begin{equation}
\sigma =\int_{x'}^{x} \sqrt{g_{ab}dx^{a}dx^{b}}=\int _{\lambda _{0}}^{\lambda} \sqrt{ g_{ab}n^{a}n^{b}}d\lambda
\label{int-rel}
\end{equation}
where $n^{a}=dx^{a}/d\lambda$ is the tangent vector to the geodesic. This equation
tells you how the metric tensor $g_{ab}$ determines the geodesic distance $\sigma^2(x,x')$. 
The second equation is the differential version of the same, given by:
 \begin{equation}
 \frac{1}{2} \left[\nabla _{a}\nabla _{b}\sigma ^{2}(x,x')\right]=g_{ab}(x) -\frac{1}{3}R_{acbd}n^cn^d\sigma^2+.... 
 =g_{ab}(x)+\mathcal{O}(R\sigma^2)
 \label{diff-rel}
  \end{equation} 
This one tells  you how $\sigma^2(x,x')$ determines the  metric tensor $g_{ab}$ when you take the limit of $x\to x'$ on both sides of this equation. (In a way  $\sigma^2(x,x')$ actually encodes more information than the metric; the  Taylor series expansion of the function $\sigma^2(x,x')$ gives the components of curvature tensor etc.) These two equations together imply that the metric $g_{ab}(x)$ and the geodesic distance
$\sigma^2(x,x')$ contain the same amount of information about the geometry. Classical gravity can, therefore,  be described entirely in terms of the single \textit{biscalar}
function $\sigma^2(x,x')$ instead of the ten components of the local metric tensor $g_{ab}(x)$.

While both $g_{ab}(x)$ and $\sigma^2(x,x')$ can be used to describe the geometry classically,  the metric is a lot easier to work with in technical computations and hence is usually considered as \textit{the} descriptor of space(time) geometry. In fact, conventional text books in general relativity hardly mention the geodesic distance!
  There are several technical reasons for this: (a) As I said before, the biscalar $\sigma^2(x,x')$ encodes more information than the metric and as such there are restrictions on what kind of functions can be accepted as the geodesic distance. For example, \eq{diff-rel} tells you that the first two coefficients of the series expansion have to be related to each other exactly as the curvature tensor is related to the metric. One can understand the existence of such restrictions from the fact that, for events which are sufficiently close together, \eq{int-rel} imposes an additive property. However, it is not easy to write down the necessary and sufficient conditions for a biscalar to be a geodesic distance. (For some discussion of this point, see e.g., \cite{rylov}) (b) There is significant algebraic complexity in working with $\sigma^2(x,x')$. For example, the Hilbert action, and the field equations, when expressed in terms of $\sigma^2(x,x')$, will involve fourth order derivatives which are very intractable. 
  
It turns out, however, that  $\sigma^2(x,x')$ is conceptually  far better suited to describe the \textit{quantum} microstructure of spacetime.\footnote{Me and my collaborators have emphasized the superiority of $\sigma^2$ over $g_{ab}$ in modeling quantum microstructure in several of our publications (see e.g.,\cite{sidmavsg,qmetric}; for different but related point-of-view, see \cite{alvarez} and references therein.). This point of view is slowly gaining some acceptance in the later works by others and I hope this trend continues!} Let me provide two important reasons for the same. 
  
  (a) 
 There is fair amount of evidence \cite{zplimp} which suggests that, at mesoscopic scales, the primary effect of quantum gravity is to change $\sigma^2$ to another function $S(\sigma^2)$ such that $S(0)\equiv L_0^2$ is finite and non-zero, signifying a zero-point-length to the spacetime.\footnote{While most of the ideas described here will will work for arbitrary $S(\sigma^2)$, I will illustrate the results for the simple choice 
$S(\sigma^2)=\sigma^2(x,y)+L_0^2$ when appropriate; on analytic continuation to a spacetime with a mostly positive signature, the zero -point-length adds to the spatial distance.} 
 Once you know how QG corrections change $\sigma^2$, one can determine --- through  \eq{int-rel} and \eq{diff-rel} --- how the metric gets corrected by QG effects. The resulting structure, called qmetric, $q_{ab}(x,x')$, is defined in analogy with \eq{diff-rel} with $\sigma^2$ replaced by $S[\sigma^2$]:
  \begin{equation}
\frac{1}{2} \left[\nabla _{a}\nabla _{b}S(x,x')\right]=q_{ab}(x,x') -\frac{1}{3}R_{acbd}n^cn^dS+.... 
 \label{expsi}
\end{equation} 
All the quantities on the right are now evaluated for the qmetric and the result holds to the lowest order in $L_0^2$. This equation defines  a bitensor $q_{ab}(x,x')$ rather than a tensor.  
  The properties of the qmetric was explored in detail in Ref. \cite{qmetric}. 
  
  (b) There is sufficient evidence to suggest that the field equations of gravity have the same conceptual status as the equations of elasticity/fluid mechanics; in other words, gravity is an emergent phenomenon (see e.g., \cite{tpreviews}). This suggests that geometrical variables should emerge from suitable, quantum mechanical, pregeometric description  of spacetime. I will suggest in this work that there is a natural way of interpreting the geodesic distance $\sigma^2(x,y)$ as a correlator of  a pregeometric \textit{density of spacetime events}, $J(x)$, in the form 
  \begin{equation}
   S[\sigma^2(x,y)] = \langle J(x) J(y)\rangle
   \label{190rev}
  \end{equation} 
  The left hand side gives the quantum corrected geodesic distance at mesoscopic scales; the probability distribution defining the correlator in the right hand side will be provided as we go along. 
  From \eq{expsi} we can also express the quantum corrected qmetric also as a correlator:
   \begin{equation}
q_{ab}(x,x')=\frac{1}{2} \langle J(x')\nabla _{a}\nabla _{b}J(x)\rangle 
 \label{expsi1}
\end{equation} 
which is valid to the lowest order in $RS$. While there is no simple way to introduce the metric tensor as an emergent construct, we can directly obtain the quantum corrected qmetric, at mesoscopic scales, as an emergent variable, if we adopt the geodesic distance as the fundamental descriptor of geometry and use \eq{190rev} and \eq{expsi1}.
  Clearly, $\sigma^2(x,y)$ is better suited for an emergent description of geometry compared to the metric tensor. 
  This is the most important reason for using $\sigma^2(x,x')$ rather than $g_{ab}(x)$ as a descriptor of geometry.\footnote{May be one could add a third reason:  It is very unlikely that  physicists would have been tempted to quantize a biscalar $\sigma^2(x,x')$! On the other hand, a second rank symmetric tensor $g_{ab}(x)$ lured them to try their luck with creating a quantum theory for the metric tensor (with   repeated failures). There is sufficient evidence that metric tensor should not be thought of as a garden-variety field (e.g. Yang-Mills field) and subjected to some kind of quantization. If general relativity was taught in terms of $\sigma^2$ may be one would have realized this earlier.}

  In other words, the \textit{classical} geodesic distance is \textit{the} descriptor of spacetime \textit{geometry} while the \textit{quantum} corrected geodesic distance $S(\sigma^2)$ can be related to the  description of \textit{pregeometric} variables through \eq{190rev}. Let me summarize this point-of-view:
  \begin{itemize}
   \item It is better to describe  geometry in terms of $\sigma^2(x,y)$ rather than the metric $g_{ab}(x)$, especially close to Planck scales.
   
   \item Spacetime geometry is an emergent phenomenon and quantum gravitational effects change $\sigma^2(x,y)$ to $S[\sigma^2(x,y)]$ with $S[0]\equiv L_0^2$ being non-zero.. 
   
   \item One can describe the geodesic distance $\sigma^2(x,y)$ and $S[\sigma^2]$ as  correlators of a pregeometric variable $J(x)$ which can be thought of as the density of spacetime events. 
   \end{itemize}
   
  \noindent I will now describe how such a picture emerges.
  
  \section{Geodesic distance emerges as a correlator of a pregeometric variable}
  
  Let me first outline the algebraic aspects and then take up the physical interpretation. 
  I start with the abstract space of all events $\{\mathcal{A}\}$ which eventually will become a set of all points in a D-dimensional Euclidean manifold with each event being identified by some coordinates $x^i$ in a local chart. One can equivalently start with a D-dimensional Euclidean manifold divided into cells of volume $L_0^D$ with the events at the centers of the cells (labeled by coordinates $x^i$) becoming the elements of the set $\{\mathcal{A}\}$. A volume $dV_x$ will correspond to  $d^Dx/L_0^D$ cells in the appropriate limit. It is convenient, though not absolutely necessary, to imagine that the Euclidean space is compact, say, a D-dimensional hypersphere of  radius $R$. (This is indeed what will happen if the Lorentzian spacetime is deSitter which becomes a hypersphere in the Euclidean sector.) We then have a finite number (about $(R/L)^D$) of elements in the set 
  $\{\mathcal{A}\}$. 
  
  I will next introduce a 
  stochastic variable $J(\mathcal{A})$, described by the probability function $\mathcal{P}[J(\mathcal{A})]$, with the leading order behaviour given by:
  \begin{align}
   \mathcal{P}[J(\mathcal{A})] &= N\exp\left[ -\frac{L_0^2}{2} \sum_{\mathcal{A},\mathcal{B}} J(\mathcal{A})S^{-1}[\sigma^2(\mathcal{A},\mathcal{B})]J(\mathcal{B}) +\cdots \right]\nonumber\\
   &= N\exp\left[ -\frac{L_0^2}{2} \sum_{x,y} J(x)S^{-1}[\sigma^2(x,y)]J(y) +\cdots \right]
   \label{191}
  \end{align} 
   which $N$ is an unimportant normalization factor which I will not display when it is irrelevant. In the second line we have used a more intuitive notation replacing elements of the set $\{\mathcal{A}\}$ by their coordinates but it must be interpreted through the first equation. In such a description, $J(\mathcal{A})\equiv J_x$ etc are a set of correlated Gaussian stochastic variables. Similarly $S[\sigma^2(\mathcal{A},\mathcal{B})]\equiv S_{(x,y)}$ is a symmetric matrix whose elements are determined by the function $\sigma^2(x,y)$ with $S^{-1}$ being the matrix inverse. So the argument of the exponential is essentially the familiar discrete matrix structure $J_xS^{-1}_{xy}J_y$.  It is then obvious that the correlator $\langle J(\mathcal{A}) J(\mathcal{B})= \rangle \langle J(x) J(y) \rangle $ is indeed given by $S[\sigma^2(x,y)]/L_0^2$. (In \eq{191}, $J$ is dimensionless and $S$ has dimensions of square of length.) In the semi-classical limit, with $x,y$ etc. are treated as coordinates of a differential manifold with the geodesic distance $\sigma^2(x,y)$ in the limit of $L_0 \to 0$. The relation in \eq{190rev} will hold even in this classical limit for events $(x,y)$ for which $\sigma^2(x,y) \gg L_0^2$. 
  
  While the above ideas work for an arbitrary $S[\sigma^2(x,y)]$, let me illustrate them in the special case in which $S[\sigma^2(x,y)] = \sigma^2(x,y) + L_0^2$ where $L_0$ is the zero point length in spacetime (which, as I said before, is taken to be of the order of the Planck length $L_P$). In this specific case \eq{191} becomes
  \begin{equation}
\mathcal{P}[J(x)] = N\exp\left[ -\frac{L_0^2}{2} \sum_{x,y} J(x)J(y) [\sigma^2(x,y) + L_0^2 ]^{-1}+\cdots \right]
   \label{191a}
  \end{equation} 
  In both \eq{191} and \eq{191a}, the $J(x)$ can be thought of as the pregeometric  \textit{density of spacetime events}. Let me now introduce this concept taking $D=4$ for definiteness.
  
  This is best done by noting that such a description (and terminology)  is completely analogous to what you do in the description of a fluid made of discrete molecules. In such a case, one often talks about a  function $n(x) = n(t, \bm{x})$ which is supposed to give the number density of molecules at an event $x^i$. Obviously an  event $x^i$ of geometrically  zero size cannot host a single molecule of finite size, let alone several of them. But you never worry about this aspect when you do fluid mechanics. This mathematical abstraction is based on the idea $dN = n(x) d^3 \bm{x}$ can describe the number of molecules in a `small' volume $d^3\bm{x}$. This volume, however, should be  large enough  to contain   sufficient number of molecules (i.e, it is significantly larger than $\lambda^3 $ where $\lambda$ is the mean free path) and hence cannot be strictly infinitesimal. At the same time, it is taken to be small enough to be treated as infinitesimal for mathematical purpose. This is precisely what I do while introducing the density of spacetime events $J(x)$; the idea is that
  \begin{equation}
   dN_e=J(x)\frac{dV_x}{L_0^4}
   \label{jden}
  \end{equation} 
 gives the number of pregeometric events around $x^i$, just as  $dN = n(x) d^3 \bm{x}$ gives the number of molecules around $x^i$.
  The argument $x^i$ of the function $J(x)$ refers to a coordinate label in a coarse grained, mesoscopic description of the spacetime, treated like a fluid, with $L_0$ being analogous to the mean-free-path. I stress that there are no conceptual ambiguities in using such a description while claiming that $S[\sigma^2(x,y)]$ is an emergent variable.

  These relations  suggest that $G(x,y) \equiv L_0^2S^{-1}(x,y)$ can be interpreted as a propagator for a theory with $J(x)$ acting \textit{as the sources}. The coincidence limit of this propagator $G(x,x) = 1$ remains finite due to the existence of the zero-point-length. Such a source $J(x)$ could also be thought of as generating a field $\phi(x)$ with  $\mathcal{P}[J(x)]$ becoming the partition function for the theory. That is, we can also write:
\begin{equation}
\mathcal{P}[J(x)]\propto \int \mathcal{D}\phi\exp
\left[-\frac{1}{2L_0^2}\sum_{x,y} \phi(x)S[\sigma^2(x,y)]\phi(x)+\sum_x J(x)\phi(x)\right]
\end{equation}
where  $G(x,y)$ is the inverse of $S(x,y)$.                                                                                                                                   
In general, the field theory for $\phi(x)$ will be  non-local. 

  In this approach, \textit{it is  important to note that I have defined  $S[\sigma^2(x,y)]$ as a correlator of the  source $J(x)$ in the discrete limit, rather than as a correlator of the field $\phi(x)$.} The correlator of a \textit{field}, in the limit of $L_P \to 0$, will usually behave as a inverse function of $\sigma^2(x,y)$; for e.g., a massless free field in $D=4$ will have a behaviour $1/\sigma^2$. But in the limit of $L_P \to 0$ we want the correlator in \eq{190rev} to behave as $\sigma^2(x,y)$. This is achieved by defining the geodesic distance in terms of the correlator of the \textit{source} in the discrete limit. That is, the geodesic distance emerges from a pregeometric variable analogous to the \textit{source} $J(x)$ rather than from a pregeometric variable analogous to the \textit{field }$\phi(x)$. It is also clear from \eq{190rev} that the correlator $\langle J(x)J(y)\rangle$ increases without bound (rather than decrease) with the separation $(x-y)^2$; obviously we are dealing with a somewhat counter-intuitive, but well-defined, stochastic variable.
  
  Since the field theory of $\phi(x)$ is difficult to handle, we need another route to reach the propagator $G(x,y)$. Fortunately, the relevant formalism for propagators with finite coincidence limit has already been developed \cite{pid} in terms of path integrals (without introducing any field). In this approach, it is
  relatively straightforward to obtain the propagator $G(x,y) \equiv (L_0^2/ S[\sigma^2(x,y)])$ directly through the introduction of zero-point length in spacetime in the continuum limit. To do this, consider a spacetime with a classical metric $g_{ab}$ and a corresponding Laplacian $\Box_g$. The heat kernel for this spacetime is defined in the standard manner as $K_{\rm std} (x,y;s) \equiv \bk{x}{e^{s\Box_g}}{y}$. The propagator for a scalar field, incorporating the zero-point length, can now be defined as
  \begin{equation}
   G(x,y) = L_0^2\int_0^\infty ds\ e^{-(L_0^2/4s)} \, K_{\rm std} (x,y; s)
   \label{193}
  \end{equation} 
  This construction has been extensively discussed in the literature \cite{zplK} and allows the modification of $\sigma^2(x,y)$ to $\sigma^2(x,y)+L_0^2$ at the lowest order. The probability distribution for the source $J(x)$ for the scalar field is given by
   \begin{equation}
 \mathcal{P}_\phi[J(x)] = N\exp\left[ -\frac{L_0^2}{2} \int_0^\infty ds e^{-(L_0^2/4s)} \sum_{x,y}\ J(x)J(y) K_{\rm std} (x,y; s)\,  +\cdots \right]
   \label{1940}
  \end{equation} 
  The probability distribution in \eq{191a}, governing the density of spacetime events, can  be expressed in a similar form. For example, when $\sigma^2(x,y)$ is a function of $(x-y)$, let $B(k)$ be the Fourier transform of $S[\sigma^2(x-y)]$ and  define the  heat kernel $K_{\rm dse} (x,y;s) \equiv \bk{x}{e^{-sB(i\partial)}}{y}$, then
  \begin{equation}
 \mathcal{P}[J(x)] = N\exp\left[ -\frac{L_0^2}{2} \int_0^\infty ds  \sum_{x,y}\ J(x)J(y) K_{\rm dse} (x,y; s)\,  +\cdots \right]
   \label{194}
  \end{equation} 
  In classical geometry the propagation amplitude from $x$ to $y$ is governed by $K_{\rm std} (x,y;s)$. In the above expression we are using a heat kernel with two modifications: 
  (1) The events $x$ and $y$ are weighted by the  density of spacetime events $J(x)$ and $J(y)$. 
  (2) The propagation at scales less than $L_0^2$ is suppressed effectively by the factor $e^{-L_0^2/4s}$. 
  
  It is possible to arrive at the modified propagator in \eq{193} from a different route as well \cite{pid}. This is done by postulating 
that the action for a relativistic particle, propagating in a spacetime, should remain invariant under the transformation $\sigma (x,y) \to L_0^2/\sigma (x,y)$. This can be achieved by defining the (Euclidean) propagator for a particle of mass $m$, propagating in a given spacetime by the sum over paths:
 \begin{equation}
  G(x,y) = \sum_\sigma \exp\left[- m \left(\sigma + \frac{L_0^2}{\sigma}\right)\right]
  \label{modact}
 \end{equation}
 It can be shown that \cite{pid} the resulting propagator incorporates the zero-point length exactly as given by \eq{193}. 
 For a massive particle, the modification in \eq{modact} has a very simple physical interpretation. 
 Recall that the action for a relativistic particle of mass $m$ is $A = - m \sigma = - \sigma/\lambda_c$ where $\sigma$ is the length of the path and $\lambda_c = \hbar/mc$ is the Compton wavelength of the particle. When we introduce gravity, it makes no sense to sum over paths with length $\sigma$ smaller than the Schwarzschild radius $R_g = Gm/c^2$ of the particle. This suggests suppressing the contribution from paths with $\sigma \lesssim R_g$ in some suitable manner. Assuming that this suppression preserves a duality symmetry under $\sigma \to 1/\sigma$, one arrives at the (unique) modification of the  action to the form $A_g = -(\sigma/\lambda_c) - (R_g/\sigma)$ which can be written in the form $A_g =  - (1/\lambda_c)[\sigma + (L_0^2/\sigma)]$ where $L_0$ is of the order of Planck length. 
  To obtain the results in \eq{194} etc., we can take the limit $m\to 0$ at the end of the computation. 
  
  So the procedure for obtaining $\sigma^2$ as an emergent variable from pregeometric density of spacetime events can be summarized as follows:  
  \begin{itemize}
   \item 
 You start with a classical spacetime and metric $g_{ab}$ valid at macroscopic scales. Compute the  heat kernel for this metric, e.g., $K_{\rm dse} (x,y;s) \equiv \bk{x}{e^{-sB(i\partial)}}{y}$. 
  
\item Postulate that the probability $\mathcal{P}[J(x)]$ for the density of spacetime events $J(x)$ is given by  \eq{194}. 
  
\item Then the quantum corrected geodesic distance $S[\sigma^2(x,y)]$ is given by the correlator $\langle J(x) J(y) \rangle$. This provides an emergent description for the geodesic distance from which one can obtain an (emergent) QG corrected metric, $q_{ab}$. The properties of such a qmetric has been explored in detail in Ref. \cite{qmetric}. 

  \end{itemize}

It should be clear that the zero-point-length plays two distinct roles in the above discussion. First, spacetime geometry acquires a zero-point-length from pregeometric variables when geodesic distance is treated as the correlator $<J(x)J(y)>$. Second, propagators for matter fields acquire finite coincidence limit [with zero-point-length acting as a UV regulator] when the path integral is modified as in  
\eq{modact}. In the complete picture, we should be able to derive 
\eq{modact} from pregeometric physics, once the description has a zero-point-length built in. This is related to the extremely nontrivial question: How do we describe \textit{matter fields} close to Planck scales? 

The \eq{modact} suggests a clue in this simple case. Note the a massive scalar field is completely specified a single parameter $m$ leading  to the standard QFT propagator 
\begin{equation}
 G(x,y;m^2)=\int_0^\infty ds e^{-m^2s}K_{std}(x,y;s)
\end{equation} 
where $K_{std}$ is the \textit{zero mass} Schwinger (heat) kernel which only depends on the geometry and is independent of any properties of the quantum field.
One can compute $G(x,y;m^2)$ --- without introducing a scalar field operator, vacuum state etc --- by directly summing over all paths connecting the points $x$ and $y$, with a weightage of $\exp- m\ell(x,y)$ for a path of length $\ell(x,y)$. Recall that this sum can be evaluated by working in a lattice with lattice separation $\epsilon$ and a corresponding mass $\mu(\epsilon)$ in the lattice. At the end of the calculation one takes the limit $\epsilon\to0$ arranging the continuum mass to be $m$. (See e.g., \cite{tpqft}). 

The existence of the quantum geometrical fluctuations introduces a zero-point-length to spacetime and modifies $G(x,y;m^2)$ to the propagator
\begin{equation}
 G_{QG}(x,y;m^2)=\int_0^\infty ds e^{-m^2s-L_0^2/4s}K_{std}(s; x,y)
 \label{g1}
\end{equation}
One can take into account the quantum geometrical fluctuations by allowing the lattice spacing --- introduced originally to compute $G(x,y;m^2)$ --- to fluctuate by a factor $\lambda$ with the amplitude $\mathcal{A}(\lambda)$. This is equivalent to assuming that there is an amplitude $\amp{m}{m_0}$  for a system with mass $m_0$ to appear as a system with mass $m=m_0/\lambda$ due to quantum spacetime fluctuations.\footnote{The fluctuation is taken to be a simple rescaling to ensure that when $m\to0$, we have $m_0\to0$.}
(The effect of pregeometric fluctuations on the matter field can only arise through superpositions  of mass $m_0$ which is the only matter-sector parameter in the problem). Therefore we expect the following relation to hold:
\begin{equation}
 G_{QG}(x,y;m^2)=\int_0^\infty dm_0 \amp{m}{m_0} G(x,y;m_0^2)
 \label{g2}
\end{equation} 
Using \eq{g1} and \eq{g2} and some Laplace transform tricks one can determine this amplitude $\amp{m}{m_0}$ to be:
\begin{equation}
\amp{m}{m_0}=-\frac{m_0L_0\theta(m_0^2-m^2)}{\sqrt{m_0^2-m^2}}J_1\left[L_0\sqrt{m_0^2-m^2}\right]
\end{equation}
for $m_0>m$. (There is a Dirac delta function contribution for $m=m_0$ which we have not displayed; see Appendix for details).
We now put $m_0=\lambda m$ and write $G$ as a path integral sum. This gives 
  \begin{equation}
  G_{\rm QG}(m) = \int_1^\infty d\lambda\ \mathcal{A}(m,\lambda)\, \sum_{\rm paths} e^{-m\lambda \ell}
  \label{z3}
  \end{equation}
  with 
  \begin{equation}
  \mathcal{A}(m,\lambda) = - \frac{\lambda(Lm)}{\sqrt{\lambda^2 -1}}\ J_1\left[m L\sqrt{\lambda^2-1}\,\right] 
  \label{z4}
  \end{equation}
  for $\lambda>1$. (Again we have not displayed a Dirac delta function contribution at $\lambda=1$; see Appendix.) This suggests the following interpretation: The presence of a mass $m$ in the space(time) induces fluctuations in the length scales changing $\ell \to \lambda \ell$ with an amplitude $C(m, \lambda)$. The correct propagator $G_{\rm QG}(m)$ has to be obtained by integrating over these fluctuations as well as the sum over paths. This leads to the correct propagator $G_{\rm QG}(m)$. 
This result hints at a possible way by which the path integral duality can arise from pregeometric fluctuations, along with zero-point-length for the spacetime.

For the sake of completeness, I mention that we can also determine a result analogous to \eq{g2} in the Lorentzian spacetime. We first define a function $G(x,y;m^2=q)\equiv G(x,y;q)$
by replacing $m^2$ by $q$ in the standard propagator and allowing $q$ to range over the whole real line. Then we find that:
\begin{equation}
  G_{QG}(x,y;m_0^2)=\int_{-\infty}^\infty\frac{dq}{2\pi} G(x,y;q)\frac{L_0}{\sqrt{q-m_0^2}}K_1\left[iL_0\sqrt{q-m_0^2}\right]
\end{equation} 
It is, however, often convenient to work in the Euclidean spacetime to avoid ambiguities in the analytic structure.

 \section{Euclidean spacetime as a set of Lorentzian null surfaces}
  
 I will now take a closer look at the conceptual picture which emerges from \eq{jden}, using the analogy of fluid mechanics.  
 Let us ask what is the operational procedure for determining the number density of molecules $n(x)$ in fluid around an event $x^i$. One simple method will be to take a small, spherical, volume $V(\epsilon)$ of radius $\epsilon$ centered at $\bf{x}$ (on a $t=$ constant  
 hypersurface) and count the number $N(\epsilon)$ of molecules in it. We can then define the number density $n(x)$ as the limit of   
 $N (\epsilon)/V(\epsilon)$ as $\epsilon\to0$. One can introduce a similar procedure to define $J(x)$, in the Euclidean space, using  a geodesic sphere of radius $\epsilon$. Interestingly enough, something very curious happens when we analytically continue to the Lorentzian spacetime. It turns out that, shifting the attention from the metric to geodesic distance, provides fresh insights into the nature of null surfaces in spacetime.
 Given the fact that null surfaces seem to play a vital role in the emergent gravity paradigm \cite{tpreviews}, let me highlight this alternative point of view.
  
  Consider a $D$-dimensional, flat, Euclidean space described in Cartesian coordinates $x^i_E = (t_E, x_E, \bp{x})$ where $\bp{x}$ denotes $(D-2)$ transverse coordinates.
  I will concentrate on the $t-x$ plane to illustrate the ideas which can be easily generalized to the $D$-dimensional space. Let us begin by asking how we can assign to a point $\mathcal{P}$ in this plane the coordinates, say, $(T_E, X_E)$. The simple procedure is just to say that the coordinates of $\mathcal{P}$ are  specified by the equations:
  \begin{equation}
   t_E =T_E; \qquad x_E = X_E
   \label{route1}
  \end{equation} 
  It is, however, possible to specify the coordinates of a point in a different --- but completely equivalent --- manner. One can say that the coordinates of $\mathcal{P}$ are given by the solution to the equation:
  \begin{equation}
  \sigma^2_E(x,X) \equiv  (x_E - X_E)^2 + (t_E - T_E)^2 =0
    \label{route2}
  \end{equation} 
  That is, we set the geodesic distance between the two points to be equal to zero.
  In the Euclidean spacetime the procedures in \eq{route1} and \eq{route2} lead to identical results, because the  Euclidean geodesic interval $\sigma_E^2(x,X)$ vanishes only if $x=X$. 
  
  Let us now analytically continue from the Euclidean space to Lorentzian spacetime by the usual procedure of setting $t_E = it$ etc. The \eq{route1} will continue to work and one can specify the Lorentzian coordinates of the event $\mathcal{P}$ by the relations $x=X$ and $t=T$. But the procedure in \eq{route2} now fails! The analytically continued version of \eq{route2} 
  \begin{equation}
  \sigma^2(x,X) \equiv  (x - X)^2 - (t - T)^2 =0
    \label{route3}
  \end{equation}
  gives the null surfaces originating at the event $(T,X)$.
  This is the direct consequence of the fact that the vanishing of the geodesic distance $\sigma^2(x,X) =0$ in a \textit{Lorentzian} spacetime specifies events connected by a null ray rather than a unique event.

  Let us take this idea further. Let us assume that the pregeometric description and consequent QG effects are to be described primarily in the Euclidean space with an analytic continuation leading to the standard Lorentzian spacetime. In that case, an infinitely localized point in the Euclidean space will not have any operational significance. To tackle this issue, we can describe a point in the Euclidean space by the following procedure. Consider the equi-geodesic surfaces defined by the equation $\sigma_E^2(x,X) = \epsilon^2$. In the context of the 2-dimensional section of Euclidean flat space this equation describes a circle of radius $\epsilon$ centered on the point $(T,X)$. If we now take smaller and smaller equigeodesic surfaces by decreasing the value of $\epsilon$ the circles will approach the events $(T,X)$. So an event in Euclidean space can equivalently be thought of as the limit of the equigeodesic surface $\sigma^2_E(x,X) = \epsilon^2$ when the geodesic distance tends to zero. this is precisely the construction we should adopt to define $J(x)$.
  
  Let us now repeat the same exercise after the analytic continuation. In the Lorentzian spacetime, the corresponding equation $\sigma^2(x,X) = \epsilon^2$ will represent a pair of hyperbola in the right and left wedges demarcated by the null surfaces $(x-X) = \pm (t-T)$. The limit $\epsilon \to 0$ will now give you the null surfaces $(x-X) = \pm (t-T)$ rather than a unique event. So if we choose to define events in Euclidean space by taking the limit $\epsilon \to 0$ of the equation $\sigma_E^2(x,X) = \epsilon^2$, then analytic continuation \textit{will associate a pair of null surfaces with each point in the Euclidean space! }
  
  This result, while algebraically elementary, has deep conceptual significance. As I described earlier, geodesic distance provides a natural link between pregeometry and geometry. It then makes sense to define the coordinates of an event in terms of the procedure in \eq{route2} in Euclidean space. On analytic continuation, this procedure associates a pair of null surfaces with pairs of points in the Euclidean space. In other words, pairs of points in the Euclidean space have a natural correspondence with null surfaces in Lorentzian spacetime. Because the construction relies only on the $\epsilon \to 0$ limit, the same ideas carry over to a curved Euclidean space and curved spacetime. In the curved Euclidean space, one can always choose a locally flat coordinate system and perform the analytic continuation within that region leading to exactly the same results. 
  
  This approach links naturally with the idea of density of spacetime events $J(x)$. One can now consider an infinitesimal region around an event $X^i$  defined through the relation $\sigma_E^2 (x,X) \le \epsilon^2$; this would represent a geodesic ball of radius $\epsilon$ in the Euclidean space and one can think of number of spacetime events  inside such  infinitesimal balls as encoded in the pregeometric variable $J(x)$.  Analytic continuation will now lead to density of spacetime events in the infinitesimal neighbourhood of a pair of null surfaces in the spacetime. 
  I would conjecture that the importance of null surfaces in the study of horizon thermodynamics and emergent gravity paradigm arises because infinitesimally localized points satisfying $\sigma_E^2(x,X) =0$ in the Euclidean space are mapped to events along the null rays, satisfying $\sigma^2(x,X) =0$ in the spacetime. Let me elaborate on this point, 
  
 Thermal behaviour in these spacetimes arise \textit{only because} we introduce singular coordinate transformations in spacetime such that, say, the lapse function vanishes on some surface. As an illustration, consider a class of \textit{Euclidean} spacetimes with line element:
  \begin{equation}
 ds^2 = F(\rho) \, [dt_E^2 + dx_E^2] + dL_\perp^2 = F(\rho)\, [\rho^2 d\theta^2 + d\rho^2] + dL_\perp^2 
     \label{14a}
   \end{equation} 
 where $\rho^2 \equiv x_E^2 + t_E^2$ and  $dL_\perp^2=h_{AB}(\rho, y^A)dy^Ady^B $ is the metric in the transverse directions.
 Consider a class of such metrics  which satisfies the following two conditions: Near the origin of the $t_E-x_E$ plane (that is when $\rho\to0$), (a) $F(\rho) \to 1$  and (b) the transverse line element $dL_\perp^2 $ reduces to the flat form $d\bp{x}^2$. This implies that the coordinate system ($t_E,x_E, \bp{x}$) becomes a locally flat, Cartesian, coordinate system near the origin. 
 In the polar form of the metric, we do have a coordinate singularity at $\rho=0$ but this is completely harmless.
 One can rewrite the line element in \eq{14a} in a more familiar form as:
   \begin{equation}
  ds^2 = f(\xi) \, d\theta^2 + \frac{d\xi^2}{f(\xi)} + dL_\perp^2
    \label{14c}
   \end{equation} 
where $f(\xi)=\rho^2 F(\rho)$ and the coordinate $\xi$ is related to $\rho$ through the relations: 
   \begin{equation}
    \rho^2 \equiv  e^{2\xi_*}; \qquad \xi_* \equiv   \int \frac{d\xi}{f(\xi)} + \mathrm{const}
    \label{defrstar} 
   \end{equation}
   Near the origin,  $\rho^2\approx 2\xi$ and the function $f(\xi)$ behaves as $f(\xi)\approx 2\xi$. This metric also has a coordinate singularity at the origin because $f$ vanishes. Again this is no more scary that the vanishing of $g_{\theta\theta}$ in the polar coordinates in \eq{14a}. 
   
   The trouble is, on analytic continuation ($\theta\to i\tau$) to Lorentzian sector, the surface $\xi=0=\rho$ becomes a 
   null surface and acts a horizon for a class of observers. The metric now takes the familiar Lorentzian form:
    \begin{equation}
  ds^2 = -f(\xi) \, d\tau^2 + \frac{d\xi^2}{f(\xi)} + dL_\perp^2
    \label{14cl}
   \end{equation} 
   This vanishing of $f$ at $\xi=0$ (which is a null surface) is now a big deal because all of horizon thermodynamics in the 
   Lorentzian version of the metric in \eq{14cl} arises from the vanishing of the lapse function $N=f$. If you study the field equation of a scalar field in these spacetimes, you will get (out-going) mode functions which behave as $\xi^{- i\omega}$ (with our scaling in which the  surface gravity is set to unity) and becomes ill-defined at $\xi=0$. The ratio of mode functions on the two sides of the horizon are related by the factor $\phi(\delta)/\phi(-\delta)=(-1)^{- i\omega}$ which is ill-defined.
  If we use the standard $i\epsilon$ prescription and the result
   \begin{equation}
    (\xi-i\epsilon)^{- i\omega}=\xi^{- i\omega}\theta(\xi) + |\xi|^{- i\omega}e^{-\pi\omega}\theta(-\xi)
   \end{equation} 
  we find that the square of the relative amplitudes of this decomposition $ e^{-2\pi\omega}$ is precisely what leads to the temperature for the null surface. Clearly the singular transformation to polar coordinates in Euclidean sector (making the origin special) leads to nontrivial effects in Lorentzian spacetime.
  
  If we try to regulate the singularity something surprising happens. To see this clearly consider a class spacetimes of the form \cite{tptopo}:
  \begin{equation}
   ds^2= f(x_E)dt_E^2+dx_E^2+dL_\perp^2\Longrightarrow -f(x)dt^2+dx^2+dL_\perp^2;\quad f(x)\equiv b^2 +a^2x^2
  \end{equation} 
  where the arrow denotes standard analytic continuation and $a,b$ are real constants. In the Euclidean sector, $a\to0, b\neq0$ limit will give flat space in Cartesian coordinates (locally) while 
  $a\neq0, b\to0$ limit will give flat space in polar coordinates (locally). In general this spacetime is curved with 
  \begin{equation}
   R=-\frac{1}{2}R^{tx}_{tx}=\frac{2a^2b^2}{(b^2 +a^2x^2)^2}
   \label{rr}
  \end{equation} 
  Looking at the numerator of \eq{rr}, you might naively think that the curvature vanishes when either $a$ or $b$ vanishes, as expected. It is indeed true that when $a\to0, b\neq0$ the curvature vanishes. But when we take $b\to0, a\neq0$, we get the limit:\footnote{This is most easily proved using the following theorem \cite{theorem}: If $F(r)$ is a unit normalized function, then the
sequence of functions $F_\mu(r) = \mu^{-1} F(r/\mu)$ tends to the Dirac delta function when $\mu\to0.$]}
  \begin{equation}
   \lim_{b\to0}R^{tx}_{tx}\propto \lim_{b\to0}\frac{b^2}{(b^2 +a^2x^2)^2}\propto\delta_D(x^2)
  \end{equation} 
  showing that the curvature is non-zero and concentrated on the origin in the Euclidean sector \textit{and on the horizon} in the Lorentzian sector! Clearly when the metric becomes flat for some limiting value for a parameter, curvature can acquire a distributional behaviour. This again reinforces the non-triviality of analytic continuation mapping the Euclidean origin to Lorentzian horizon.

  \section{Summary and Highlights}
  
 Let me conclude by listing  some of the key points described above. 
  
  \begin{itemize}
   \item The geodesic distance is a far better descriptor of classical geometry and quantum pregeometry. Let us abandon the description in terms of the metric and concentrate on the geodesic interval,  in the study of quantum spacetime! 
   \item The geodesic interval $\sigma^2(x,y)$ is an emergent construct and can be thought of as a correlator $\langle J(x)J(y) \rangle$ of a pregeometric variable $J(x)$. This variable, which can be interpreted as the density of spacetime events,  is completely analogous to density of molecules $n(x)$ of a fluid in standard fluid mechanics.
   \item The correlator  $\langle J(x)J(y) \rangle$ is computed using a probability functional given by \eq{194}. This provides a systematic procedure for constructing the quantum corrected metric from the classical metric. 
   \item  Events in Euclidean space can be defined by taking the  zero radius  limit  ($\epsilon \to 0$) of a geodesic ball $\sigma^2_E(x,X) = \epsilon^2$. The analytic continuation of this exercise allows us to associate a pair of null surfaces with every point in Euclidean plane. 
   \item  It is possible to introduce a pregeometric field $\phi(x)$ sourced by the density of spacetime events $J(x)$ by the procedure I have outlined.  For reasons I have stressed, the geodesic interval should be thought of as the correlator of the \textit{source} $J(x)$ and \textit{not} of the field $\phi(x)$.
   \item  One can investigate the pregeometric structure as well as specific examples (cosmological spacetimes, black hole spacetimes etc.) working with either $J(x)$ or $\phi(x)$. Standard Lorentz invariant, local, unitary QFT will not allow a propagator with finite coincidence limit. So the QFT of $\phi(x)$ will be non-local and rather unusual. 
  \end{itemize}

  \section*{Acknowledgements}
  
I thank Sumanta Chakraborty and Dawood Kothawala for  discussions.  My research is partially
supported by the J.C.Bose Fellowship of Department of Science and Technology, Government of India.

\section*{Appendix: Calculational Details}

This Appendix provides some background material on zero point length and the algebraic details for some of the results quoted in the main text. 

\subsection*{Propagator with zero point length}

The standard Euclidean propagator  $G_{\rm std}$ can in flat space(time) be obtained by the path integral sum 
  \begin{equation}
   G_{\rm std} (x,y; m^2) = \sum_{\rm paths} \exp\left[ -m \ell\right]
  \end{equation} 
  where $\ell(x,y) $ is the length of a path connecting the points $x$ and $y$ and the sum is over all paths: This sum can be defined by a lattice regularization procedure (see e.g., \cite{tpqft}) thereby leading to the standard result. (The amplitude $e^{-m\ell}$ is well defined for $m>0$ and is divergent for $m<0$. However, the lattice regularization leads to a result for $G(p)$ which depends only on $m^2$. This invariance of the result under $m\to -m$ is a peculiar feature of the measure used in the lattice to regularize the sum.)
  The introduction of the zero point length into the  space(time) suggests modifying the path integral sum \cite{pid} to the form
  \begin{equation}
   G_{\rm QG} (x,y) = \sum_{\rm paths} \exp\left[ -m\left(\ell + \frac{L^2}{4\ell}\right)\right]\, ; \qquad L^2 = \mathcal{O}(1)\, L_P^2
  \end{equation} 
  In flat space(time), this path integral sum can also be worked out by lattice regularization techniques \cite{pid} and the final result can be expressed in the form:
  \begin{equation}
   G_{\rm QG}(p^2) = \int_0^\infty ds \ \exp[-s(p^2+m^2) - \frac{L^2}{4s}] = \frac{L}{\sqrt{p^2+m^2}} \ K_1[L\sqrt{p^2+m^2}]
  \end{equation} 
  Incidentally, the \textit{same} propagator $G_{\rm QG}(p^2)$ can also be obtained by the integral
  \begin{equation}
   G(p^2) = \int_0^\infty d\mu \ \exp\left[-\frac{\mu }{G_{\rm QG}(p^2)}\right] \equiv \int_0^\infty d\mu \ \exp[-\mu F(p^2)]\, ; \qquad F(p^2) \equiv \frac{1}{G_{\rm QG}(p^2)}
  \end{equation} 
  where I have defined $F(p^2) = 1/G_{\rm QG}(p^2)$. It follows that the propagator in real space can be expressed in a form: 
  \begin{equation}
   G_{\rm QG} (x,y; m^2) = \int_0^\infty d\mu \ \bk{x}{e^{-\mu F(-\Box)}}{y}
  \end{equation} 
  So $G_{\rm QG}(x,y)$ can be thought of as a propagator for a field theory with a  non-polynomial Lagrangian leading to the action
  \begin{equation}
   A = \int d^Dx\ \phi(x) F(-\Box)\, \phi(x)
  \end{equation} 
  Unfortunately, this field theory is difficult to handle.
  
  In a curved space(time),  the zero point length is introduced in a similar manner, by modifying the relation between heat kernel and the propagator into the form:
 \begin{align}
 G_{\rm QG}(x_1,x_2;m^2) &= \int_0^\infty ds\ e^{-m^2 s - \frac{L^2}{4s}} \, K_{m=0}( x_2, x_1; s)
  \end{align}
  The corresponding expression in the Lorentzian spacetime is given by: 
  \begin{align}
 G_{\rm QG}(x_1,x_2;m^2) &= \int_0^\infty ds\ e^{-im^2 s +\frac{iL^2}{4s}} \, K_{m=0}( x_2, x_1; s)
 \label{newten}
\end{align}
 The sign of $L^2$  term reflects the fact that zero point length adds to \textit{spatial} part of $\sigma^2(x,y)$. With the signature being mostly negative, this requires $\sigma^2 \to \sigma^2 - L^2$. 
 
 \subsection*{Higher dimensional interpretation of $G_{\rm QG}$}
 
 It is  possible to express $G_{\rm QG}$ in $D$-dimensions in terms of the \textit{standard} propagator in $N=D+2$ dimensions. To do this, consider a fictitious $N=D+2$, Euclidean \textit{curved} space(time) with the metric
  \begin{equation}
  dS_N^2 = \left(g_{ab} dx^a dx^b\right)_D + \delta_{AB}\ dX^A dX^B  \qquad \qquad (A,B = 1,2)
   \label{y1}
  \end{equation} 
  where we have added two ``flat'' directions. The $N$-dimensional heat kernel for $m=0$ now factorizes and we can write  
\begin{equation}
 K^N_{m=0} \equiv \bk{x,\bm{L}}{e^{s\Box_N}}{y,\bm{0}}
 =\left(\frac{1}{4\pi s}\right) e^{-L^2/4s} \, K^D_{m=0}(x,y; s) \, ;\qquad L^2 \equiv L_AL^A
   \label{y2}
  \end{equation} 
  Therefore, the $N$-dimensional, massive, heat kernel becomes  
\begin{equation}
 K^N(x,\bm{L}; y,\bm{0}; s) = \left(\frac{1}{4\pi s}\right) e^{-(L^2/4s) - m^2s} \, K^D_{m=0}(x,y; s)
   \label{y3}
  \end{equation} 
  The corresponding $N$-dimensional massive propagator is obtained by integrating this expression over $s$ in the range 0 to $\infty$. This is almost the same as $G_{\rm QG}$ except for the extra factor $(1/4\pi s)$.  This factor can be easily taken care of by Differentiating the propagator with respect to $m^2$. We find that  
\begin{equation}
 -4\pi \frac{\partial}{\partial m^2} \, G_{\rm std}^N(x,\bm{L}; y,\bm{0}) = \int_0^\infty ds\ e^{-(L^2/4s) - m^2s} \, K^D_{m=0}(x,y; s)
  = G_{\rm QG}^D (x,y) 
  \end{equation}
  Therefore, we can relate the quantum corrected propagator $G_{\rm QG}^D(x,y)$ to the standard propagator in the fictitious $N=D+2$ space with the metric in \eq{y1} by the relation
  \begin{equation}
   G_{\rm QG}^D(x,y) = - 4\pi \frac{\partial}{\partial m^2} G^N(x,\bm{L}; y, \bm{0}) \bigg|_{\mathbf{L}^2=L^2}
    \label{y4}
  \end{equation} 
 The zero point length arises as the magnitude of the propagation distance in the extra dimensions. 
  
  Interestingly, the derivative of standard propagator with respect to $m^2$ --- which appears in the right hand side of \eq{y4} ---can be related to the `transitivity integral' for the propagator $G^N$. To see this, consider the integral  (with the notation  $d^N\bar z \equiv d^Dx\, d^2x$): 
  \begin{align}
  \int d^N\bar z \, G^N(x,\bm{L}; z,\bm{Z}) \, G^N(z,\bm{Z}; y,\bm{0}) &= \int d^N\bar z \ \bk{x,L}{(-\Box^N + m^2)^{-1}}{z,Z} \bk{z,Z}{(-\Box^N + m^2)^{-1}}{y,0}\nonumber\\
  &= \bk{x,\bm{L}}{(-\Box^N + m^2)^{-2}}{y,\bm{0}} 
   \label{y5}
  \end{align}
  Using the fact that:
   \begin{equation}
   \bk{x,\bm{L}}{(-\Box^N + m^2)^{-2}}{y,\bm{0}} = -  \frac{\partial}{\partial m^2} \, G^N(x,\bm{L}; y,\bm{0})
   \label{y6}
  \end{equation}
   and combining with \eq{y4} we find that 
  \begin{equation}
  G^D_{QG} = (4\pi) \bk{x,\bm{L}}{(-\Box^N + m^2)^{-2}}{y,\bm{0}} \bigg|_{\mathbf{L}^2=L^2}
   \label{y7}
  \end{equation}
  This result, in turn, leads to two \textit{differential} equations satisfied by $G^D_{QG}$, viz.: 
  \begin{equation}
  (-\Box^N + m^2)^{2} G^D_{QG} =4\pi \delta(x,y)\, \delta(\bm{L}) 
   \label{y8}
  \end{equation}
  and 
  \begin{equation}
   (-\Box^N + m^2) G^D_{QG} = 4\pi G^N(x, \bm{L}; y,\bm{0})
   \label{y9}
  \end{equation}
  which are valid in any curved space(time).
  
  One can explicitly verify that, in flat space(time) either of these equations lead to the correct $G_{\rm QG}$. For example, Fourier transforming either \eq{y8} or \eq{y9}, we find that $G^D_{\rm QG}$ can be expressed as the integral
  \begin{equation}
  G^D_{QG} (x, \bm{L}; 0,\bm{0}) = 4\pi \int\frac{d^Dk \, d^2K}{(2\pi)^N} \ \frac{e^{ikx}\, e^{i\bm{K\cdot L}}}{(k^2 + m^2 + K^2)^2}
   \label{y10}
  \end{equation}
  The square in the denominator can be taken care of by the usual trick of differentiating with respect to $m^2$. Performing the 2-dimensional integral over $d^2K = K dK d\theta$ we get the result in terms of the Bessel function $J_0(KL)$:
  \begin{align}
  G^D_{QG} (x, \bm{L}; 0,\bm{0}) &= -\frac{\partial}{\partial m^2} \int \frac{d^Dk}{(2\pi)^D} \int_0^\infty K dK\int_0^{2\pi}\frac{d\theta}{\pi} \,  \frac{e^{ikx}\, e^{i{K L\cos\theta}}}{k^2 + m^2 + K^2} \nonumber\\
  &= -\frac{\partial}{\partial m^2} \int \frac{d^Dk}{(2\pi)^D}\, e^{ikx}\int_0^\infty 2K dK\ \frac{J_0(KL)}{k^2 + m^2 + K^2} 
   \label{y11}
  \end{align}
  This tells us that the $D$-dimensional Fourier transform  $G^D_{\rm QG} (k,\bm{L})$ of 
  $G^D_{\rm QG}(x, \bm{L}; 0,\bm{0})$  is given by 
  \begin{equation}
   G^D_{\rm QG} (k,\bm{L}) = - \frac{\partial}{\partial m^2} \int_0^\infty 2K dK \ \frac{J_0(KL)}{k^2 + m^2+K^2}
  \end{equation}
  To do the integral over $K$ we write the denominator in the exponential form and obtain 
  \begin{equation}
 -\frac{\partial}{\partial m^2} \int_0^\infty 2K dK \ J_0(KL) \int_0^\infty  ds\ e^{-s(k^2+m^2)} \, e^{-sK^2}
 = \int_0^\infty ds\ s e^{-s(k^2+m^2)} \int_0^\infty 2K \, dK\ J_0(KL) e^{-sK^2}
   \label{y12}
  \end{equation}
  We now use the  identity 
  \begin{equation}
   \int_0^\infty 2K dK \ J_0(KL) e^{-sK^2} = \frac{1}{2} \exp\left( -\frac{L^2}{4s}\right)
   \label{ythirty}
  \end{equation} 
  to recover the standard result
  \begin{equation}
  G^D_{\rm QG} (k,{L}) = \int_0^\infty ds\ e^{-s(k^2+m^2) - (L^2/4s)}
   \label{y13}
  \end{equation}

\subsection*{Determining $G_{\rm QG}$ in terms of $G_{\rm std}$}
 
 \subsubsection*{Lorentzian spacetime}

  It is often convenient to express the modified propagator $G_{\rm QG}(x_1,x_2;m^2)$ directly in terms of the original propagator $G_{\rm std}(x_1,x_2;m^2)$. This is fairly straightforward to do in the Lorentzian sector along the following lines. We  set $\mu \equiv m^2$ and treat $G_{\rm std}(x_1,x_2;m^2)=G_{\rm std}(x_1,x_2;\mu)$ as a function of $\mu$ analytically continued along the entire real line. The relation between the propagator and the zero mass heat kernel
  \begin{equation}
 G_{\rm std}(x_1,x_2;\mu) = \int_0^\infty ds\ e^{-i\mu s} \, K_{m=0} ( x_2, x_1; s)
  \end{equation}
  can now be inverted to give:
  \begin{equation}
 \int_{-\infty}^\infty G(x_1,x_2;\mu)\, e^{i\mu q} \frac{d\mu}{2\pi} = \int_0^\infty ds \ \delta(q-s)\, K = \theta(q) K_{m=0} (x_2, x_1; q)
  \end{equation}
  We can therefore write \eq{newten} as:
  \begin{align}
 G_{\rm QG}(x_1,x_2;m^2) &= \int_0^\infty ds\ e^{-im^2 s +(iL^2/4s)} \, K_{m=0}( x_2, x_1; s)\nonumber\\ 
 &= \int_0^\infty ds\  e^{-im^2 s +(iL^2/4s)} \int_{-\infty}^\infty \frac{d\mu}{2\pi} \ e^{i\mu s}G_{\rm std}(x_1,x_2;\mu)\nonumber\\
 &= \int_{-\infty}^\infty \frac{d\mu}{2\pi} \, G_{\rm std}(x_1,x_2;\mu) \int_0^\infty ds\, e^{-is(m^2 -\mu) +(iL^2/4s)}
  \end{align}
  The integral  over $s$ can be evaluated in terms of $K_1$ giving
  \begin{equation}
  \int_0^\infty ds\, e^{-is(m^2 -\mu) +(iL^2/4s)}= \frac{L}{\sqrt{\mu - m^2}} \, K_1 (iL\sqrt{\mu - m^2})
  \end{equation} 
  This leads to the result
  \begin{equation}
 G_{\rm QG}(x_1,x_2;m^2) = \int_{-\infty}^\infty \frac{d\mu}{2\pi} \ G_{\rm std} (x_1,x_2; m^2=\mu) \ \frac{L}{\sqrt{\mu -m^2}}\, K_1(iL\sqrt{\mu - m^2})
 \label{keya4}
  \end{equation}
  which gives the modified propagator in any curved spacetime in terms of the original propagator. 
  
If the background spacetime has certain symmetries, they reflect on the structure of both $G_{\rm std}$ and  $G_{\rm QG}$ in a similar manner. For example, consider homogeneous spacetimes like the FRW spacetimes in which  $G_{\rm std} (x_1,x_2; m^2)=
 G_{\rm std} (\mathbf{x}_1-\mathbf{x}_2, t_1,t_2; m^2) $. It follows that, we will also have $G_{\rm QG} (x_1,x_2; m^2)=
 G_{\rm QG} (\mathbf{x}_1-\mathbf{x}_2, t_1,t_2; m^2) $. This implies one can Fourier transform both propagators with respect to spatial coordinates and obtain a similar relation in the momentum space:
 \begin{equation}
 G_{\rm QG}(\mathbf{p}, t_1,t_2;m^2) = \int_{-\infty}^\infty \frac{d\mu}{2\pi} \ G_{\rm std} (\mathbf{p}, t_1,t_2; m^2=\mu) \ \frac{L}{\sqrt{\mu -m^2}}\, K_1(iL\sqrt{\mu - m^2})
  \end{equation}
on the other hand, if the spacetime is static, one can Fourier transform both propagators with respect to time coordinates and obtain a similar relation:  
 \begin{equation}
 G_{\rm QG}(\omega, \mathbf{x}_1,\mathbf{x}_2;m^2) = \int_{-\infty}^\infty \frac{d\mu}{2\pi} \ G_{\rm std} (\omega, \mathbf{x}_1,\mathbf{x}_2; m^2=\mu) \ \frac{L}{\sqrt{\mu -m^2}}\, K_1(iL\sqrt{\mu - m^2})
  \end{equation}
 More complicated symmetries can be handles in a similar manner. For example, in maximally symmetric spacetimes, both propagators will (essentially) depend on the geodesic distance and one can often deal with the Fourier transform with respect to the geodesic distance. In specific cases one can also express $G_{\rm std}$ in terms of the mode functions and obtain a modified set of mode function corresponding to $G_{\rm QG}$.
 
 It is easy to verify that, in the case of flat spacetime we do reproduce the correct result from \eq{keya4}. 
  In Fourier space $G_{\rm std} (p^2,m^2 = \mu)$ is given by
  \begin{equation}
   G_{\rm std}(p^2,\mu) = \frac{i}{p^2 -\mu + i\epsilon} = -  \frac{i}{\mu -p^2 - i\epsilon}
  \end{equation} 
  which has a pole at $\mu = p^2 + i\epsilon$. It is obvious from the properties of $K_1$ that the integral can be evaluated by closing the contour  in the upper half of the complex plane for $s>0$. The contour then contributes $2\pi i$ times the residue at  the pole at $\mu = p^2 + i\epsilon$ thereby leading to the correct propagator.

\subsubsection*{Euclidean space}

Here we need to use inverse Laplace transform rather than inverse Fourier transform. This is most easily done by
starting with the integral:
  \begin{align}
 \int_a^\infty dt \ e^{-pt} J_0\left[ 2 \sqrt{b(t-a)}\right] = \frac{1}{p} e^{-ap - (b/p)}
   \label{1mar} 
  \end{align} 
  which can be verified by setting $b(t-a) = x^2 $ and using \eq{ythirty}.
  Differentiating both sides of \eq{1mar} with respect to $a$ we get
  \begin{equation}
 e^{-ap - (b/p)} =  - \frac{\partial}{\partial a} \int_a^\infty dt\ e^{-pt} J_0\left[2 \sqrt{b(t-a)}\right]
   \label{two}
  \end{equation} 
  The integration limits in the right hand side can be extended from 0 to $\infty$ by introducing a factor $\theta[t-a]$ in the integrand.
    Carrying  the differential operator $\partial/\partial a$ inside the integral, one will then obtain one term containing $\theta J_1$ as well as another term of the form $J_0 \delta$ giving rise to $e^{-ap}$. That is:
     \begin{align}
  - e^{-ap - (b/p)} &= \frac{\partial}{\partial a} \int_a^\infty e^{-pt}\, J_0\left[ 2 \sqrt{b(t-a)}\right] dt  
  = - e^{-ap} + \int_a^\infty dt \ e^{-pt}\ b\ \frac{J_1\left[ 2 \sqrt{b(t-a)}\right]}{\sqrt{b(t-a)}} \nonumber\\
  &= \int_0^\infty dt \ e^{-pt} \left\{ \theta(t-a) \frac{b}{\sqrt{b(t-a)}} \, J_1 \left( 2\sqrt{b(t-a)}\right) - \delta(t-a)\right\}
   \label{2mar} 
  \end{align}    
    It is however more convenient \textit{not} to do this and instead use the expression in \eq{two} as it is in the computations. The differentiation can be carried out towards the end, when required. We will now set $a=m^2$ and $b=L^2/4$ in \eq{two} to obtain: 
  \begin{equation}
  e^{-m^2s - (L^2/4s)} =  - \frac{\partial}{\partial m^2} \int_{m^2}^\infty dt\ e^{-st} J_0\left[L \sqrt{t-m^2}\right]
  =- \frac{\partial}{\partial m^2}\int_{m^2}^\infty d m_0^2\ e^{-m_0^2s} J_0\left[L \sqrt{m_0^2-m^2}\right]
   \label{three}
  \end{equation} 
  where, in the second step, we have put $t=m_0^2$. We will now use this expansion in the definition of quantum gravitational propagator given by
  \begin{equation}
   G_{\rm QG} (m^2) = \int_0^\infty ds\ e^{-m^2s - (L^2/4s)} K_0(s)
   \label{four}
  \end{equation} 
  Here $K_0(s) = \bk{x}{e^{s\Box}}{y}$ is the zero mass heat kernel in an arbitrary curved space(time)\footnote{I have suppressed the dependence of $K_0$ and $G_{\rm QG}$ on the coordinates $x,y$. When the metric is independent of some of the coordinates, the same relation can be used in momentum space as well because the integrals for Fourier transform just flow through the expressions in both sides.}  with $\Box \equiv \Box_g$ being the Laplacian corresponding to the curved space metric $g_{ab}$.
  Using \eq{three} in \eq{four} we find that
  \begin{align}
  G_{\rm QG}(m^2) &= - \int_0^\infty ds \ K_0(s) \frac{\partial}{\partial m^2} \int_{m^2}^\infty dm_0^2 \ e^{-m_0^2 s} J_0\left[L \sqrt{m_0^2-m^2}\right]\nonumber\\
  &= -\frac{\partial}{\partial m^2} \int_{m^2}^\infty dm_0^2 \ J_0\left[L \sqrt{m_0^2-m^2}\right] \int_0^\infty ds \ K_0(s) e^{-m^2_0 s}\nonumber\\
  &= - \frac{\partial}{\partial m^2} \int_{m^2}^\infty dm_0^2 \ J_0\left[L \sqrt{m_0^2-m^2}\right] G_{\rm std} (m_0)
   \label{five}
  \end{align} 
  In arriving at the last equality we have used the fact that the integral over $s$ gives the standard QFT propagator (without quantum corrections) corresponding to a mass $m_0$.
  Equation (\ref{five}) directly relates the quantum corrected propagator for mass $m$ to the standard QFT propagator for mass $m_0$ in an arbitrary Euclidean space(time). Whenever the latter is known, the former can be computed.  This is the Euclidean version of \eq{keya4}. The comments related to the symmetries of the propagator etc., made in the context of Lorentzian spacetime, continue to be valid in this case as well.
  
 It is easy to verify this result for the flat space(time) in which $G_{\rm std}$ in momentum space is given by 
  \begin{equation}
   G_{\rm std}(p^2,m_0^2) = \int_0^\infty d\mu\ e^{-\mu(p^2 + m_0^2)}
   \label{six}
  \end{equation} 
  Using this expression in \eq{five}, changing variable to  $x^2 \equiv m_0^2 -m^2$ and carrying out the integrals, we find that
  \begin{align}
    G_{\rm QG}(m^2) & = -  \frac{\partial}{\partial m^2} \int_0^\infty 2 x \ dx\ J_0(Lx) \int_0^\infty d\mu\ e^{-\mu p^2} \, e^{-\mu(m^2 + x^2)}\nonumber\\
    & = -  \frac{\partial}{\partial m^2} \int_0^\infty d\mu\ e^{-\mu(p^2 + m^2)} \int_0^\infty 2 x \ dx\ J_0(Lx)e^{-\mu x^2} 
    =  \int_0^\infty d\mu\ e^{-\mu(p^2 + m^2)- (L^2/4\mu)}
   \label{seven}
  \end{align} 
  where, to obtain the last equality, we have used the  identity in \eq{ythirty}.
  Clearly, \eq{seven} gives the correct quantum gravitational propagator in flat space(time). 
  
This result in \eq{five} allows us to write $G_{\rm QG}$ for a mass $m$ as a convolution of the standard propagator $G_{\rm std}(m_0)$ for mass $m_0$ through the relation
  \begin{equation}
 G_{\rm QG}(m) = \int_0^\infty dm_0\ C(m,m_0)\, G_{\rm std}(m_0)
   \label{z1}
  \end{equation}
  where 
  \begin{equation}
 C(m,m_0) = - \frac{m_0L\theta(m^2_0-m^2)}{\sqrt{m^2_0-m^2}}\, J_1\left[L\sqrt{m^2_0-m^2}\,\right] + \delta(m-m_0)
   \label{z2}
  \end{equation}
  and we have used \eq{2mar}.
  We can now put $m_0=\lambda m$ and write $G_{\rm std}$ as a path integral sum, which will lead to \eq{z3} and \eq{z4} in the main text.


\begin{thebibliography}{000}

\bibitem{synge} 

J.L. Synge, \textit{Relativity: the general theory} (North-Holland, Amsterdam, 1960).

\bibitem{rylov}

Y.A. Rylov (1990), Jour. Math. Physics, \textbf{31}, 2876


\bibitem{sidmavsg} 
  
  Kothawala D and Padmanabhan T (2014) \textit{Phys. Rev.} \textbf{D 90} 124060 [arXiv:1405.4967];
  
  Kothawala D (2013) \textit{Phys. Rev.} \textbf{D 88} 104029
  
  Padmanabhan, T (2015) \textit{Entropy}, \textbf{17},  7420 [arXiv:1508.06286]
  
  D. Jaffino Stargen, D. Kothawala, (2015), \textit{Phys.Rev.}, \textbf{D92}  024046 [arXiv:1503.03793]
  
  \bibitem{zplimp}
  B. S. DeWitt, (1964) \textit{Phys. Rev. Lett.} \textbf{13}, 114; 
  
  T.Padmanabhan, (1985), \textit{Gen. Rel. Grav.,}  \textbf{17}, 215; 
  
  T.Padmanabhan, (1985), \textit{Ann. Phys.},  \textbf{165}, 38; 
  
  T.Padmanabhan, (1987), \textit{Class. Quan. Grav.},  \textbf{4}, L107;
  
  For a review, see L. Garay, \textit{Int. J. Mod. Phys.} \textbf{A 10}, 145 (1995); S. Hossenfelder, \textit{Living Rev. Relativity} \textbf{16}, (2013), 2  [arXiv:1203.6191]
  
  \bibitem{qmetric} 
  Padmanabhan T, Chakraborty S and Kothawala D (2016) \textit{Gen. Rel. Grav.}, \textbf{48} 55 [arXiv:1507.05669];
  
  Kothawala D and Padmanabhan T (2015) \textit{Phys. Lett.} \textbf{B 748}, 67;
  
  Sumanta Chakraborty, D. Kothawala, Alessandro Pesci, (2019) \textit{Phys. Lett.} \textbf{B 797}, 134877  [arXiv:1904.09053];
  
  Alessandro Pesci, \textit{Class. Quantum Grav.,} \textbf{36} (2019) 075009 [arXiv:1812.01275]; 
  
  Alessandro Pesci, \textit{Looking at spacetime atoms from within the Lorentz sector}, [arXiv:1803.05726]
  
  \bibitem{alvarez} 
  
  Alvarez, E. et al. \textit{Phys.Rev.}, \textbf{D45} (1992) 2033 
  
  \bibitem{tpreviews}
  
   T. Padmanabhan,  \textit{Gen.Rel.Grav,} \textbf{46}, 1673 (2014) [arXiv:1312.3253];
   
   T. Padmanabhan, \textit{Gravity and Quantum Theory: Domains of Conflict and Contact,} IJMPD, 2030001 (2020) [arXiv:1909.02015]

  

  \bibitem{zplK} 
  K.Srinivasan, L.Sriramkumar and T. Padmanabhan,  (1998) \textit{Phys. Rev.} \textbf{D 58}, 044009  [gr-qc-9710104];
  
  S. Shankaranarayanan and T. Padmanabhan,  (2001) \textit{Int. Jour. Mod. Phys }, \textbf{10 },  351 [gr-qc-0003058];
  
  Dawood Kothawala, L. Sriramkumar, S. Shankaranarayanan, T. Padmanabhan, (2009), \textit{Phys.Rev.,} \textbf{D 79}, 104020  [arXiv:0904.3217] 
  
  \bibitem{pid} 
  
  T. Padmanabhan,  \textit{Phys. Rev. Letts}, \textbf{78}, 1854 (1997) [hep-th-9608182]; 
  
  T. Padmanabhan,  \textit{Phys. Rev.}, \textbf{D 57} , 6206 (1998) 
  
  \bibitem{tpqft}
  T. Padmanabhan, (2016) \textit{Quantum Field Theory: The Why, What and How} Springer, Heidelberg 
  
  
  \bibitem{tptopo}
  T. Padmanabhan, \textit{Mod. Phys. Letts}, \textbf{A18}, 2903 (2003) [hep-th/0302068] 
  
  \bibitem{theorem}
  Stakgold, Ivar. (1979), \textit{Green’s functions and boundary value problems,} John Wiley, New York
 p.110.
  
  \end{thebibliography}
  \end{document}